# Bilevel optimisation with embedded neural networks: Application to scheduling and control integration.

Roberto X. Jiménez J.

Supervisor: Dr. Antonio del Rio-Chanona
Co-supervisor: Damien van de Berg

Submitted to the Faculty of Engineering
Department of Chemical Engineering
Imperial College London

**September 2022**

**This is the preprint of my M.Sc. thesis**

# Bilevel optimisation with embedded neural networks: Application to scheduling and control integration.

Roberto X. Jiménez J.[a],

[a] Siemens Process Systems Engineering Limited, London, UK

**Abstract:**

Decision-making architectures in the process industries require scheduling problems to explicitly account for control considerations in their optimisation. The literature proposes two traditional ways to solve this integrated problem: hierarchical and monolithic approaches. The monolithic approach ignores the control level's objective and incorporates it as a constraint into the upper level at the cost of suboptimality. The hierarchical approach requires solving a mathematically complex bilevel problem with the scheduling acting as the leader/upper-level and control as the follower/lower level. The linking variables between both levels belong to a small subset of scheduling and control decision variables. For this subset of variables, data-driven surrogate models have been used to learn follower responses to different leader decisions. In this work, we propose to use ReLU neural networks for the control level. Consequently, the bilevel problem is collapsed into a single-level MILP that is still able to account for the control level's objective. This single-level MILP reformulation is compared with the monolithic approach and benchmarked against embedding a nonlinear expression of the neural networks into the optimisation. Moreover, a neural network is used to predict control level feasibility. The case studies involve batch reactor and sequential batch process scheduling problems. The proposed methodology finds optimal solutions while largely outperforming both approaches in terms of computational time. Additionally, due to well-developed MILP solvers, adding ReLU neural networks in a MILP form marginally impacts the computational time. The solution's error due to prediction accuracy is correlated with the neural network training error and can lead to suboptimal solutions. Overall, we expose how - by using an existing big-M reformulation and being careful about integrating machine learning and optimisation pipelines - we can more efficiently solve the bilevel scheduling-control problem with high accuracy.

## 1. Introduction

Sequential decision-making processes in hierarchical industrial operations (planning, scheduling, and control) lead to suboptimal or infeasible solutions (Dias & Ierapetritou, 2017). Therefore, integrated approaches are required to account for the lower-level limitations in upper-level decisions (Baldea & Harjunkoski, 2014). This coordination allows for more flexible and profitable operations. Coordinated decisions between scheduling and control are essential to guarantee dynamically feasible operations (Daoutidis et al., 2018).

Scheduling problems belong to a class of mixed-integer-linear programming (MILP) problems in which, for their worst-case scenario, the computational time grows exponentially with the model size (NP-hard) (Floudas & Lin, 2004). On the other hand, optimal control or dynamic optimisation problems involve a system of differential-algebraic equations, which are discretised to be solved as nonlinear programming (NLP) problems. Therefore, integrating scheduling and control would result in a large and high-dimensional MINLP. Besides the substantial computational effort required to solve the resulting MINLP, scheduling and control integration is also hindered due to conflicting objectives and different time scales (Li & Ierapetritou, 2008). For instance, the scheduling could pursue an economic objective while the control could focus on a safety-related objective. Moreover, the control level time scale typically varies from seconds to hours, while the scheduling timeframe ranges from days to weeks.

To solve the integrated scheduling and control, two methods have been pursued in literature: the monolithic/centralised approach and the hierarchical approach. Studies have focused on the monolithic approach, which aims to obtain dynamically aware schedules by integrating the control-level equations as constraints into the scheduling (Tsay & Baldea, 2020; Dias et al., 2018; Burnak et al., 2018; Caspari et al., 2020; Zhuge & Ierapetritou, 2012, 2014). However, this is done at the expense of ignoring the control level's objective. The resulting MINLPs are solved by decomposition algorithms (Generalized Benders Decomposition or Outer-Approximation algorithms). These algorithms require solving a MILP (master problem) and an NLP (primal problem).

At the same time, few studies have explored a hierarchical approach, which poses the integration as a bilevel optimisation problem with the scheduling as the leader/upper-level and the dynamic optimisation as the follower/lower-level (Chu & You, 2014; Beykal, Avraamidou & Pistikopoulos, 2022). The bilevel optimisation problem could be solved by methods such as the Karush-Kuhn-Tucker single-level reduction, decent methods, stochastic approaches and bilevel decomposition (Sinha, Malo & Deb, 2018). Nevertheless, it is well-known that bilevel optimisation problems are hard to solve, even for simple cases.

In general, linear approximations and reformulations of nonlinear terms have been used to transform MINLPs to MILPs at the expense of adding new variables or compromising the solution's accuracy (Grossmann, 2012). In integrated scheduling and control, nonlinearities arise from the control level. Therefore, the control level could be approximated by a linear surrogate model to speed up the solution and integrated into the mixed-integer-linear scheduling level (Tsay & Baldea, 2019).

Surrogate models such as decision trees and Rectified Linear Unit (ReLU) neural networks have been successfully formulated as MILP and embedded in optimisation problems (Thebelt et al., 2021; Mišić, 2020; Biggs & Hariss, 2017; Schweidtmann & Mitsos, 2019; Tsay et al., 2021). Neural networks are well-known due to their capability to approximate nonlinear functions, while decision trees are good at handling discrete outputs. However, embedding surrogate models in optimisation problems can lead to constraint violations due to approximation errors. The literature has already explored integrated operations decision-making via data-driven surrogate models. For example, Sachio et al. (2021) proposed the integration of design and control using reinforcement learning, and Dias & Ierapetritou (2019) evaluated different data-driven surrogate models' performance to solve the planning and scheduling integrated problem.

This work aims to integrate scheduling and control to enable coordinated decisions while respecting the autonomy of both levels. The bilevel optimisation problem will be reformulated as a single-level problem by replacing the control level with a data-driven surrogate model (e.g., a ReLU neural network). This data-driven surrogate model can be posed as a MILP and seamlessly embedded into the already MILP scheduling problem. Furthermore, since the surrogate model formulation does not change the type of optimisation, the resulting problem could be solved faster than the MINLP problems from other approaches. Finally, this work shows how careful integration of optimisation and machine learning pipelines could drastically reduce mathematical complexity and computational effort.

## 2. Background

This section will cover the concepts required to understand the proposed methodology for solving bilevel scheduling-control optimisation problems with embedded data-driven surrogate models. It starts with a brief overview of each problem separately (scheduling and dynamic optimisation), followed by an integrated scheduling and control methods review. Finally, an overview of bilevel optimisation with embedded data-driven surrogate models emphasising neural networks is given.

### 2.1. Dynamic Optimisation

The control level is posed as a dynamic optimisation/optimal control problem. Equations 2.1 describe a typical dynamic optimisation problem in chemical engineering, which aims to determine the optimal control trajectories *u(t)*, time invariants **v** or final time $t_f$ to minimise an objective function *J* while satisfying equalities (2.1b) and inequalities (2.1c) constraints. The set of equalities constraints in (2.1b) is a system of differential and algebraic equations (DAE) with its respective set of initial conditions **h**. The DAE system describes a mechanistic model from a chemical process/unit operation, where **x** is the vector of differential or state variables, and **y** are algebraic variables.

$$\min_{\mathbf{u}(t),\mathbf{v},t_f} J(\mathbf{x}(t),\mathbf{y}(t),\mathbf{u}(t),\mathbf{v}) \tag{2.1a}$$

$$\begin{aligned}&\mathbf{f}(\dot{\mathbf{x}}(t),\mathbf{x}(t),\mathbf{y}(t),\mathbf{u}(t),\mathbf{v}) = 0\\&\mathbf{g}(\mathbf{x}(t),\mathbf{y}(t),\mathbf{u}(t),\mathbf{v}) = 0\\&\mathbf{h}(\mathbf{x}(t),\mathbf{y}(t),\mathbf{v}) = 0\end{aligned} \tag{2.1b}$$

$$\boldsymbol{\psi}(\mathbf{x}(t),\mathbf{y}(t),\mathbf{u}(t),\mathbf{v}) \leq \boldsymbol{\psi}^U \tag{2.1c}$$

$$\mathbf{v} \in [v^L, v^U], \quad \mathbf{u}(t) \in [u^L, u^U] \tag{2.1d}$$

$$t_0 \leq t \leq t_f \tag{2.1e}$$

$$\mathbf{x} \in \mathbb{R}^{n_x}, \mathbf{y} \in \mathbb{R}^{n_y}, \mathbf{v} \in \mathbb{R}^{n_v}, \mathbf{u} \in \mathbb{R}^{n_u}$$
$$J: \mathbb{R}^{n_x \cdot n_y \cdot n_u \cdot n_v} \longrightarrow \mathbb{R}^{n_J}$$
$$\mathbf{f}: \mathbb{R}^{n_x \cdot n_y \cdot n_u \cdot n_v} \longrightarrow \mathbb{R}^{n_f}, \mathbf{g}: \mathbb{R}^{n_x \cdot n_y \cdot n_u \cdot n_v} \longrightarrow \mathbb{R}^{n_g},$$
$$\mathbf{h}: \mathbb{R}^{n_x \cdot n_y \cdot n_u} \longrightarrow \mathbb{R}^{n_h}, \boldsymbol{\psi}: \mathbb{R}^{n_x \cdot n_y \cdot n_u \cdot n_v} \longrightarrow \mathbb{R}^{n_\psi}$$

Dynamic optimisation problems are reformulated as NLPs by discretising the differential equations and parametrising the control variables. The resulting problem is nonlinear and high-dimensional. A standard discretisation scheme is the Lagrange-Radau collocation method, which approximates functions to Lagrange polynomials and enforces them to match the actual function values at given points (roots of the Radau polynomial) to guarantee good approximations; for more information, the reader is directed to Nicholson et al. (2018). Consequently, a larger discretisation will approximate functions better; however, it will lead to larger NLPs and require more computational effort.

### 2.2. Scheduling of network represented processes

Scheduling deals with resource allocation (raw material, equipment, utilities, etc.) and time sequencing decisions to minimise cost or makespan while satisfying production targets within a given time horizon (Schilling & Pantelides, 1996; Pinto & Grossmann, 1998). Therefore, it is a combinatorial problem in nature. In general, scheduling algorithms are task-centred and have been classified according to how the process is represented (resource task network and state task network) and time representations (continuous and discrete) (Floudas & Lin, 2004, 2005; Giménez, Henning & Maravelias, 2009; Harjunkoski et al., 2014). A task defines an operation (e.g., reaction) to be performed in a unit (reactor) using determined resources (raw material and utilities) (Castro, Grossmann & Zhang, 2018).

Regarding the process representation, the Resource Task Network (Figure 2.1a) considers raw materials, units, utilities, and intermediates products as resources, which explicitly allows expressing resource assignment as flow balance constraints (Nie et al., 2014; Wassick & Ferrio, 2011; Giménez, Henning & Maravelias, 2009). On the other hand, the State Task Network (Figure 2.1b) only represents connections between materials (states) produced/consumed and tasks. States' consumption and production are formulated as material balance constraints. Unit assignment and utility consumption are mapped and modelled by assignment/combinatorial constraints (Giménez, Henning & Maravelias, 2009; Maravelias, 2005; Maravelias & Grossmann, 2003; Ierapetritou & Floudas, 1998).

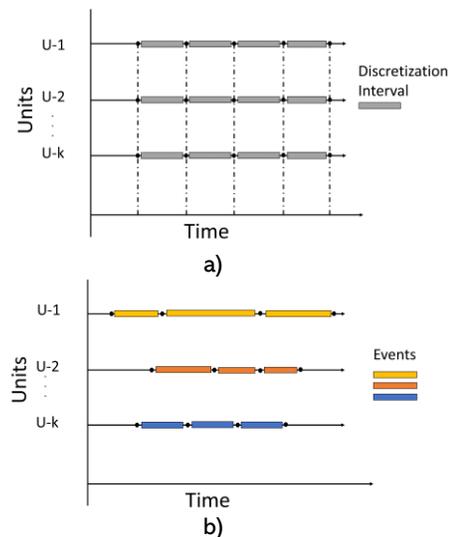

Figure 2.2. Time representation in scheduling algorithms. a) Discrete-time, b) Continuous-time.

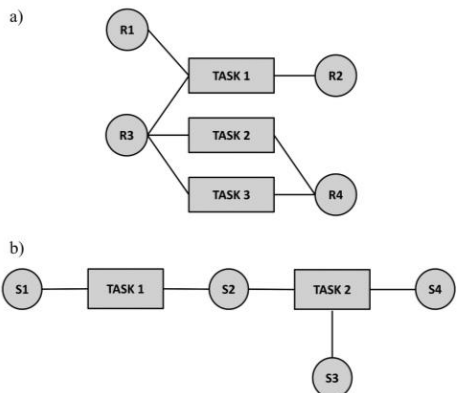

Figure 2.1. Network process representation for scheduling algorithms. a) Resource Task Network, b) State task Network.

Figure 2.2 shows the two typical time representations encountered in literature. Figure 2.2a depicts a regular grid discretisation of the scheduling time horizon. As can be seen, the task's beginning and ending points are restricted to discrete values. Therefore, the scheduling problem's size inversely increases with the discretisation intervals, which is usually equal to the shortest processing time among all tasks. In a continuous time representation (Figure 2.2b), the idea of events is introduced. An event determines the start and end of a task, and a finite number of events must be determined a priori. Therefore, a task can start and end at any moment since time is modelled as a continuous variable. (Floudas & Lin, 2005; Lee & Maravelias, 2018; Nie et al., 2014).

For the purpose of formulating the scheduling-control bilevel optimisation problems, a proper scheduling model that allows straight communication with the lower level must be selected. For instance, the State Task Network allows keeping the standard process flowsheet variables' structure. At the same time, a continuous time representation would be able to accommodate any processing time returned from the control level.

### 2.3. Integration of Scheduling and Control

Ideally, scheduling and control must always be solved in a simultaneous manner. However, it is most likely to be sequentially performed. For instance, the scheduling is individually optimised to determine production targets, and its outputs are passed to the dynamic optimisation problem. The dynamic optimisation uses this information as setpoints to determine the optimal control trajectories. Consequently, this approach could lead to suboptimal or infeasible solutions because dynamic limitations from the control level were not considered while making scheduling decisions. The papers reviewed below have proved that integrated approaches outperform the sequential approach in terms of optimal solution at the expense of adding high mathematical complexity. Two scheduling and control integrated approaches are encountered in literature: centralised/monolithic and the hierarchical model (Chu & You, 2015). Overall, the main problem for both integrated methodologies is the computational burden due to the need to solve the large MINLP. Figure 2.3 illustrates the differences between both methods.

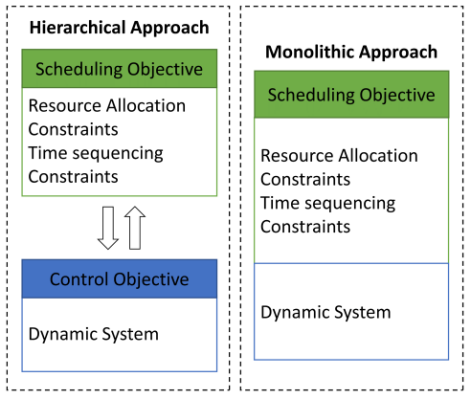

Figure 2.3. Integration of scheduling and control representation (Monolithic and hierarchical approaches).

The monolithic approach has been widely explored in literature. Most work has focused on replacing the control level with a low-order dynamic model built from systematic model reduction techniques or system identification (Tsay & Baldea, 2020; Daoutidis et al., 2018). Literature has explored the demand-respond scheduling of continuous processes. For instance, Caspari et al. (2020) solved the scheduling of an air separation unit to minimise the operational cost in a variable electricity price scenario. The control level was replaced by a low-order dynamic model of the process dynamics and control. In a similar work, Tsay & Baldea (2020) proposed a methodology to identify scheduling relevant variables from the control level via manifold learning. After identifying the relevant variables, a low-order dynamic model was constructed and incorporated into the scheduling. Nyström, Harjunkoski & Kroll (2006) and Chu & You (2013) decoupled the integrated problem into master (scheduling) and primal (control) problems following the logic of MINLP algorithms. Shi & You (2015) replaced the control level with an adaptative piece-wise linear surrogate model and formulated the scheduling-control problem as a MILP. However, the MILP solver needed to handle a special ordered set of variables to build relaxations for the piece-wise linear surrogate model.

For the hierarchical approach, Chu & You (2014) proposed a bilevel Stackelberg game formulation. The bilevel problem was solved by a decomposition algorithm, in which the lower-level responses to the upper-level decisions were modelled by a piece-wise linear function. Overall, the main drawback of the monolithic approach is the completely lost of lower-level autonomy (Moore & Bard, 1990), while the bilevel problem in the hierarchical approach is even harder to solve (Molan & Schimdt, 2022).

In general, literature has focused on developing complex algorithms to solve the monolithic and hierarchical approaches rather than attempting to reduce the computational burden due to the recurrent evaluations of the NLP raised from the control level.

Moreover, methods that follow the leader-follower logic while avoiding complicated bilevel formulations must be explored more. For instance, Avraamidou & Pistikopoulos (2019) reformulated the bilevel scheduling and design problem as a single-level problem by parametrising the upper level with lower-level decisions.

Due to the well-developed theory in systems identification and model reduction techniques, there is a clear preference for replacing the control level with low-order dynamic models. A few papers have explored machine learning surrogate models and integrated operations optimisation pipelines. For instance, Dias & Ierapetritou (2019) used neural networks, decision trees and support vector machine (SVM) classifiers to map feasible schedules and solve the integrated planning and scheduling problem. A nonlinear complex algebraic expression was obtained for neural networks and SVM; however, obtaining algebraic expressions for large decision trees presented some complications. The expressions were incorporated into the scheduling level, and the resulting MINLP was solved. Despite being modelled as a nonlinear expression, neural networks presented the most significant reduction in computational time. The triple integration of planning, scheduling and control via machine learning models has also been explored (Dias & Ierapetritou, 2020; Chu et al., 2015). The methodology proposed to replace two levels with one surrogate model. However, replacing two levels with a single surrogate model might require a large amount of data to capture their entire behaviour. Other examples of operations integration with machine learning surrogate models can be found in Beykal, Avraamidou & Pistikopoulos (2022) and Sachio et al. (2021).

Overall, integrating operations via machine learning models seems promissory. Smart optimisation and machine learning pipelines could obtain fast results within an acceptable error. For instance, to select suitable surrogates that add little computational effort to optimisers. However, proposing a general framework for scheduling and control might be challenging due to the different mathematical formulations used at each level. This model's heterogenicity needs to be evaluated on a case basis to determine the best methodology to embed the surrogate model. Although the nonlinear expression derived from the neural network transformed the scheduling problem into an MINLP, they have outperformed other nonlinear surrogate models in terms of solution speed. Consequently, embedding a neural network while retaining the MILP scheduling structure would further reduce computational effort.

## 2.4. Optimisation with embedded neural networks.

As shown in Figure 2.4, A feed-forward neural network is an empirical model of $N_L$ layers (input, hidden and output layers) formed by $N_n$ number of nodes, which are interconnected with each other (Himmelblau, 2000). The inputs of the neural network are contained in vector **x** and mapped to $z_0$ by the input layer (Eq.2.2a). A node $n$ in layer $l$ receives an input vector $z_{l-1}$, whose size is equal to the number of nodes from layer $l-1$. The nodes' outputs $z_l$ are calculated by performing the operation in Eq. (2.2b), where $\sigma$ is the activation function. The activation function's argument is a linear combination of the input vector $z_{l-1}$, with $\omega_l$ and $b_l$ known as the weight and bias (Schweidtmann & Mitsos, 2019). The outputs of the neural network **y** are the outputs from each node of the last layer $z_{N_L}$. Activation functions can be classified as smooth (sigmoid, hyperbolic tangent, etc.) or non-smooth (e.g., ReLU). Neural networks can approximate most functions as long enough units-neurons are available (Hornik, Stinchcombe & White, 1989). Like any other supervised machine learning model, they must be trained, and their prediction must be validated and tested to ensure a good fit.

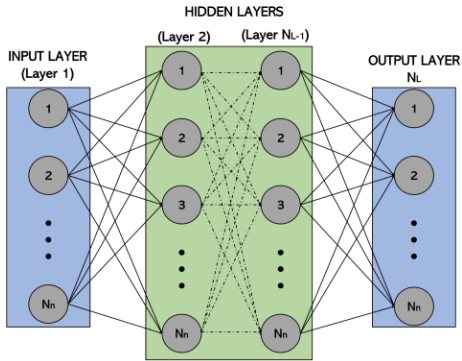

Figure 2.4. Neural network representation

$$z_0 = x \quad (2.2a)$$
$$z_l = \sigma(\omega_l^T z_{l-1} + b_l) \quad (2.2b)$$
$$y = z_{N_L} \quad (2.2c)$$
$$\sigma: \mathbb{R}^{z_{l-1}} \to \mathbb{R}^{n_{z_l}},$$

Once the neural network is trained, it can be used to estimate extreme cases (minimisation/maximisation) of the learned function. In general, two approaches are stated in the literature to embed data-driven surrogate models into optimisation problems: 1) reduced-space formulation and 2) full-space formulation (Schweidtmann, Bongartz & Mitsos, 2022). The main difference relies on the surrogate model's variables that the optimiser can manipulate. Moreover, the properties of the resulting optimisation problem (e.g., MILP, MINLP or NLP) will depend on the surrogate characteristics.

The full-space formulation could be analogously explained as an equation-oriented process model optimisation. For this class of problems, the optimiser treats the design and dependent variables as degrees of freedom, while the model equations and limitations are introduced as constraints (Bongartz & Mitsos, 2017). For neural networks, the design variables are equivalent to the inputs of the first layer. Neurons are expressed as constraints, and the dependent variables are the neuron's outputs and the neural network's final outputs. Therefore, the number of variables and constraints scales with the size of the neural network. However, simpler expressions are obtained for the constraints, and more efficient optimisation formulations that exploit linear activation functions could be developed to approximate and optimise nonlinear functions.

For example, ReLU neural networks have been formulated as MILP via big-M constraints (Anderson et al., 2018; Tsay et al., 2021; Grimstad & Andersson, 2019). Therefore, a nonlinear function can be approximated by a ReLU neural network and optimised using MILP solvers. Since MILPs solvers are branch-bound algorithms that build relaxations by relaxing binary variables, the resulting problem can be solved using well-developed linear programming solvers. However, the full-space reformulation could hinder global optimisation algorithms' convergence since convex relaxations for all the extra constraints would be needed. Furthermore, since big-M reformulation is used, tight variable bounds are required to build compact convex relaxations. These tight bounds could be hard to determine for dependent variables within a neural network.

In the reduced-space formulation, the trained neural network can be seen as an input-output function **y** = **f(x)**. Each layer's symbolic expression can be substituted as an input for the next layer. Consequently, this formulation produces large and complex algebraic expressions and cannot handle non-smooth activation functions. On the other hand, the number of variables is considerably reduced (Schweidtmann & Mitsos, 2019).

Schweidtmann & Mitsos (2019) used the reduced-space formulation to propose a global optimisation framework for problems with embedded smooth neural networks. If global optimality via spatial branch and reduce methods is not the objective, the linearity obtained from the full-space reformulation is an advantage in computational effort. For example, in integrated scheduling and control, the control level can be replaced by a ReLU neural network and embedded via the full-space formulation to the scheduling (already MILP). Moreover, MILP solvers implement heuristics rules to avoid local optimums.

Finally, few papers have been published regarding bilevel optimisation and embedded surrogate models, and two different approaches can be identified. The first reformulates the problem as a single-level optimisation problem by replacing the lower level with a surrogate model. The surrogate model is trained from historical leader-follower responses (Molan & Schimdt, 2022; Sachio et al., 2021; Dias & Ierapetritou, 2020). The second approach consists of a grey box optimisation of a surrogate model constructed from sampling the upper-level objective function for different optimal responses of the lower level (Beykal et al., 2020).

In this work, we propose to reformulate the bilevel scheduling and control problem as a single-level problem with an embedded neural network. Since not all variables from the control level are relevant to the scheduling problem, a neural network could learn optimal solutions from the dynamic optimisation problem within this reduced-dimension space. ReLU neural networks would be used to replace the control level to reduce the complexity caused by its nonlinearities and high dimensionality. The inputs of the neural network would be decision variables from the scheduling level that impact the control level, while the outputs would be control variables that affect the scheduling. Since the neural network is trained with optimal lower-level solutions, the control level would retain its autonomy. The ReLU neural network will be formulated as a MILP via full-space reformulation and seamlessly integrated into the already mixed-integer-linear scheduling problem.

This methodology takes advantage of the single-level formulation from the monolithic approach while still considering the lower-level objective, like the hierarchical approach. Moreover, a neural network will also be trained to predict feasibility. From a mathematical perspective, the objective of this work is to reduce the computational effort of a class of bilevel optimisation problems, in which the upper level is a discrete decision problem posed as a MILP while the lower level can exhibit any other complex structure (non-convex, nonlinear, dynamic system, MINLP etc.) but limited to continuous linking variables between both levels.

## 3. Methodology

Overall, the methodology starts with identifying how the upper and lower level communicates to determine linking variables. Then, data is generated from solving the lower level for different instances, and ReLU neural networks are trained, validated, and tested. Next, the trained neural networks are formulated as MILP and embedded within the upper level. Finally, the new single-level problem is solved using state-of-the-art MILP solvers. Figure 3.1 summarises the proposed methodology. This methodology is implemented using Pyomo 6.2, a python-based algebraic modelling language (Hart et al., 2011). MILPs, NLPs and MINLPs were solved using Gurobi 9.5.1, Ipopt 3.14.8 and Bonmin 1.8.7 solvers, respectively.

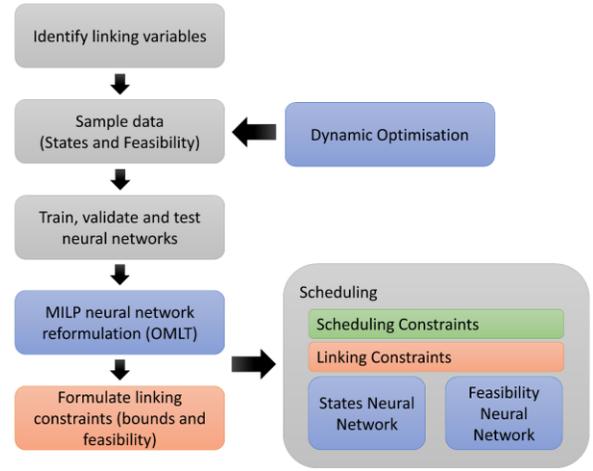

Figure 3.1. Proposed methodology.

### 3.1. Identifying linking variables

Consider the generic bilevel scheduling-control problem in Eqs. (3.1), where $\mathbf{x}$ and $\mathbf{y}$ are the scheduling and control decision variables, respectively. The linking variables $\mathbf{x_s}$ and $\mathbf{y_c}$ are subsets of $\mathbf{x}$ and $\mathbf{y}$, respectively. Subscript $s$ is used for the scheduling variables and $c$ for the control level variables. Variables $\mathbf{x_s}$ are scheduling variables relevant to the control level (neural network inputs). In contrast, $\mathbf{y_c}$ are the control level variables pertinent to the scheduling level (neural network outputs). $f$, $h$ and $g$ are the objective function, equality, and inequality constraints, respectively. A limitation of this methodology is that $\mathbf{y_s}$ must be continuous variables, as explained in the following sections. In principle, $\mathbf{x_c}$ could contain continuous or discrete variables; however, discrete input variables are not explored in this work.

$$\begin{aligned} \min_{\mathbf{x},\mathbf{y_s}} \quad & f_s(\mathbf{x},\mathbf{y_c}) \\ s.t \quad & \mathbf{h_s}(\mathbf{x},\mathbf{y_c}) = 0 \\ & \mathbf{g_s}(\mathbf{x},\mathbf{y_c}) \leq 0 \\ \min_{\mathbf{y}} \quad & f_c(\mathbf{y}) \\ s.t \quad & \mathbf{h_c}(\mathbf{x_s},\mathbf{y}) = 0 \\ & \mathbf{g_c}(\mathbf{x_s},\mathbf{y}) \leq 0 \\ \mathbf{y_c} \subseteq \mathbf{y}, \mathbf{x_s} \subseteq \mathbf{x}, & \\ \mathbf{x} \in \mathbb{R}^{n_x}, \mathbf{y} \in \mathbb{R}^{n_y} & \end{aligned} \quad (3.1)$$

In non-integrated scheduling problems, variables $\mathbf{y_c}$ are known fixed parameters (e.g., recipes), which are usually multiplied by a binary variable. This binary variable determines if a task is being performed or not at a given time. Therefore, an extra reformulation is needed to avoid nonlinearities when treated as variables in integrated scheduling and control problems.

## 3.2. Sampling, training, and validating neural networks.

Once the linking variables are identified, the control level is posed as a dynamic optimisation problem, as described in Section 2.1. Variables $x_s$ are treated as known fixed parameters. Then a regular mesh of sampling points is built within the bounds of $x_s$. The dynamic optimisation problem is solved for each mesh point, and the output variables $y_c$ are stored. Moreover, an extra output $y_f$ is added to map feasibility, and it is assigned a value of 1 (feasible) or -1 (infeasible). The dynamic optimisation problem is solved with a large enough number of discretisation points to minimise its impact on the solutions. Finally, this step could be omitted if historical data is available.

The generated dataset is used to train two regression neural networks: the states neural network and the feasibility neural network. The states neural network predicts the values of $y_c$, while the feasibility neural network is used to determine whether the control level is feasible or not for the given input values. The feasibility neural network could also be modelled as a classification problem. However, this would force the inclusion of a nonlinear activation function in the output layer. Therefore, it would not be possible to reformulate it as a MILP. This restriction to regression neural networks also limits $y_c$ to continuous variables. Since the feasibility neural network is modelled using a regression neural network, $y_f$ is a continuous variable within the range [-1,1]. Therefore, these outputs can be interpreted considering positive values for feasible points and negative values for infeasible points. Both neural networks use the ReLU activation function for the hidden layers and a linear activation function for the output layer.

The dataset is split into training, validation, and test sets. The neural networks' hyperparameters and architecture are manually tuned on a case basis to minimise the mean-squared error (MSE) for the states neural network and accuracy for the feasibility neural network, both in the test set. To calculate the feasibility neural network's accuracy, a post-processing step is required to assign -1 to negative outputs and 1 to positive ones. The learning curves from the training and validation are monitored to evaluate model learning and generalising capabilities. Neural networks are trained using TensorFlow 2.8

## 3.3. MILP Neural Network reformulation

A single rectified linear unit (Eq. 3.2a) can be interpreted as a MILP via big-M reformulation (Tsay et al., 2021; Anderson et al., 2018) as described in Equations (3.2b-3.2f), where $\varepsilon$ is a binary variable which determines whether the neuron is active or not. $M^L$ and $M^U$ are the big-M coefficients and the upper and lower bounds for the neuron's output.

$$z = \max(0, \omega^T z_{l-1} + b) \qquad (3.2a)$$
$$z \geq (\omega^T z_{l-1} + b) \qquad (3.2b)$$
$$z \leq (\omega^T z_{l-1} + b) - (1-\varepsilon)M^L \qquad (3.2c)$$
$$z \leq \varepsilon M^U \qquad (3.2d)$$
$$z \geq 0 \qquad (3.2e)$$
$$(\omega^T z_{l-1} + b) \in [M^L, M^U], \varepsilon \in \{0,1\}^{n_\varepsilon} \qquad (3.2f)$$

If $\varepsilon$ is set to 1, the neuron's output is equal to $(\omega^T z_{l-1}+b)$ due to constraints 3.2b and 3.2c. On the other hand, if $\varepsilon$ is set to 0, constraint 3.2c is relaxed, while constraints 3.2d and 3.2e force z to be 0. Notice that in the former situation, constraint 3.2b is satisfied since $(\omega^T z_{l-1}+b)$ would be negative as well as its lower bound. The outputs of the neurons from the last layer $z_{N_L}$ are variables $y_c$. Each ReLU neuron is replaced by four linear constraints and requires one binary variable. This task was automatically performed for all the neurons of the neural networks using the Optimisation and Machine Learning Toolkit 1.0 (OMLT) developed by Ceccon et al. (2022). OMLT translates a trained neural network to an algebraic language compatible with Pyomo. This procedure is applied for both states and feasibility neural networks.

## 3.4. Linking constraints

As mentioned in section 3.1, $y_c$ variables could be involved in bilinear terms. A general form for this bilinear term can be seen in Eq. 3.3a., where $x_1$ and $x_2$ are continuous upper-level variables and $q$ a binary upper-level variable. The reformulation introduces an intermediate variable $\hat{y}_c$ and the constraints shown in Eqs. 3.b-f. After adding the constraints, Eq. 3.3a is replaced by Eq. 3.3f. This reformulation automatically forces $\hat{y}_c$ to be 0 if $q$ is set 0, and the original bilinear term $(y_c \cdot q)$ is avoided.

$$x_1 = x_0 + y_c q \qquad (3.3a)$$
$$\hat{y}_c \leq Y_c^U q \qquad (3.3b)$$
$$\hat{y}_c \geq Y_c^L q \qquad (3.3c)$$
$$\hat{y}_c \leq y_c + Y_c^U(1-q) \qquad (3.3d)$$
$$\hat{y}_c \geq y_c - Y_c^L(1-q) \qquad (3.3e)$$
$$x_1 = x_0 + \hat{y}_c \qquad (3.3f)$$
$$y_c \in [Y_c^U, Y_c^L]; q \in \{0,1\} \qquad (3.3g)$$

Moreover, since a negative output of the feasibility neural network represents infeasibility from the control level, $y_f$ is directly used, as shown in Eq. 3.4. After adding these constraints, the resulting single-level problem is solved using a state-of-the-art MILP solver (Gurobi 9.5.1).

$$y_f > 0 \qquad (3.4)$$

The proposed problems are also solved via the monolithic and the reduced-space formulation (translating the neural network to a nonlinear expression). This allows evaluation of the computational effort reduction and solution error. For the rest of the text, embedding a neural network using the reduced space formulation is identified as NN-MINLP Integration. The proposed methodology would be called NN-MILP Integration.

## 4. Toy Problem

To highlight the differences between the hierarchical, monolithic, and NN-MILP integration approaches, the toy problem described in Eqs. (4.1) is solved with the tree methodologies. Two scenarios were evaluated, 1) Aligned objectives, in which both leader and follower seek the minimisation of -y, and 2) Adversarial objectives, in which the lower-level pursues the maximisation of -y while the upper-level remains the same.

$$\begin{aligned} \min_{x} \quad & x - 4y \\ s.t \quad & x \geq 0 \\ & (1) \min_{y} \; -y \; or \; (2) \max_{y} \; -y \\ & s.t \quad -x - 2y + 3 \leq 0 \\ & \qquad -2 + y \leq 0 \\ & \qquad -3x + 2y + 4 \leq 0 \\ & \qquad -12 + x + y \leq 0 \end{aligned} \quad (4.1)$$

For the hierarchical approach, the toy problem is reduced to a single-level optimisation problem by replacing the lower level with its Karush-Kunh-Tucker (KKT) optimal conditions. In the monolithic approach, the lower-level constraints are included in the upper level while its objective is ignored. The NN-MILP integration follows the methodology described in Section 3. A ReLU neural network with one hidden layer with four neurons is trained for each scenario. The neural networks obtained a mean squared error (MSE) of 1x10-4 for the test set. Their training and validation curves can be found in Figure A.1 in Appendix A. The sampled regions for the training can be seen in Figure 4.1, which agree with the respective lower-level objective for each scenario. The optimum solutions for the three methodologies can be found in Table 1.

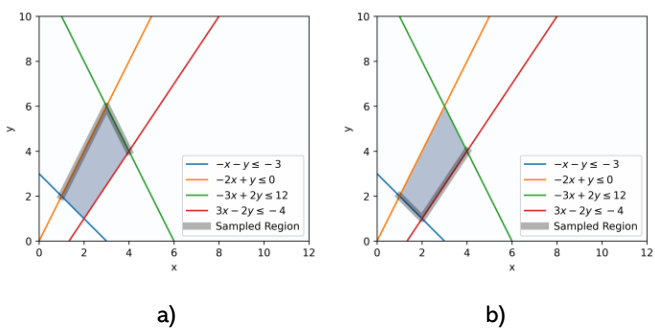

Figure 4.1. Sampled regions. a) Aligned objectives, b) Adversarial objectives

The monolithic approach finds the exact solution for the aligned scenario while being incapable of finding the correct solution for adversarial objectives. On the other hand, the NN-MILP Integration solution captures the lower-level objective but drifts from the actual value due to the training error. The addition of binary variables and constraints might be seen as unfavourable for the toy problem. However, the advantages of the neural network MILP reformulation will be elucidated in more extensive problems.

Overall, three major observations can be taken from evaluating the toy problem solution: 1) The neural network training error's impact on the solution, 2) The capability of the NN-MILP Integration to retain information from the lower-level objective, and 3) The extra constraints and binary variables needed the reformulate a neural network as a MILP.

Table 1. Toy problem solutions via different approaches. (C) Continuous variable, (B) Binary variables.

| Scenario | Hierarchical | Monolithic | NN-MILP Integration |
|---|---|---|---|
| Aligned | Obj = -21<br>x=3<br>y = 6 | Obj = -21<br>x = 3<br>y = 6 | Obj = -20.9835<br>x = 2.9982<br>y = 5.9954 |
| Adversarial | Obj = 0<br>x = 4<br>y = 4 | Obj = -21<br>x = 3<br>y = 6 | Obj = 0.0065<br>x = 4<br>y = 3.9935 |
| Extra variables/ constraints | - | Variables = 2(C)<br>Constraints = 6 | Variables = 17(C), 4(B)<br>Constraints = 29 |

## 5. Case Study 1

This case study aims to evaluate the performance of the NN-MILP Integration against the monolithic approach and NN-MINLP Integration in the scheduling of the sequential batch process presented in Figure 5.1a.

### 5.1 Problem Description

The process involves a mixer, batch reactor, and separator. First, raw materials are blended and homogenised in the mixer. Then, the batch reactor carries out a sequential reaction A→B→C, where A is the feed, B is the desired product, and C is a by-product. Finally, the separator split the reactor outlet to recover product B. The continuous-time formulation for the state task network algorithm developed by Ierapetritou & Floudas (1998) is implemented, and its entire formulation can be found in Appendix B. The scheduling level pursues the objective to maximise profit and could be summarised in Eqs. 5.1.

$$\begin{aligned} \max \quad & Profit \; (revenue - operational \; cost) \\ s.t \quad & Allocation \; Constraints \\ & Capacity \; Constraints \\ & State \; Material \; Balances \\ & Demand \; Constraints \\ & Duration \; Constraints \\ & Sequence \; Constraints \\ & Time \; Horizon \; Constraints \end{aligned} \quad (5.1)$$

The dynamic optimisation problem for the batch reactor pursues the optimal trade-off between processing time and utility consumption. The complete formulation can be found in Eqs. (5.2). The problem is an adaptation from Mishra et al. (2005). The mixer and separation processing times are proportional to the processed quantity; therefore, no dynamics are implemented for them. Moreover, their utility consumption is not considered.

$$\min_{u(t),t_f} \quad 1.2t_f + 0.5Q$$
$$s.t \quad \dot{C}_A = -u(t)C_A(t)$$
$$\dot{C}_B = u(t)C_A(t) - \beta u(t)^\alpha C_B(t);$$
$$\dot{C}_C = \beta u(t)^\alpha C_B(t);$$
$$Q = V^2 \int_0^{t_f} u(t)dt \quad (5.2)$$
$$1 \le u(t) \le 9 \ [uU.m^{-3}h^{-1}]$$
$$C_A(t=0) = 17 \ mol/m^3$$
$$C_B(t=0) = 0; C_B(t=t_f) = 13 \ [mol/m^3]$$
$$C_C(t=0) = 0; C_C(t=t_f) = 2.5 \ [mol/m^3]$$
$$\beta = 0.065; \alpha = 0.8$$

Figure 5.1b depicts how the scheduling and control level communicates when formulated as a bilevel problem. As can be seen, not all variables from the control level are relevant for the scheduling. Once the scheduling has decided the reactor batch size $V$ for a given event, the control level must calculate the optimal processing time $t_f$ and utility consumption $Q$. These three variables are the input and outputs of the state neural network.

Data is sampled by solving the dynamic optimisation problem for different batch sizes. Figure 5.2 depicts an example of optimal control solution for the batch reactor. The processing time could be reduced if more utility is consumed; however, the control level objective seeks the optimal trade-off between these two variables.

The NN-MILP Integration requires a ReLU activation function, while any smooth activation function could be used for the NN-MINLP Integration. The sigmoid activation function (Eq. 5.3) is used for the former. For comparison, both neural networks have the same architecture with two hidden layers and five neurons at each layer. The neural network predictions for the test set can be seen in Figure 5.3, and their learning curves can be found in Figure A.2 in Appendix A. The MSEs obtained for the test set are 0.0016 and 0.0036 for the ReLU and sigmoid neural networks, respectively. Since the bounds of the reactor batch size at the scheduling level are known [0, 7.5 m³] at the scheduling level, the dynamic optimisation problem can obtain feasible solutions for any batch size within the range. Therefore, the feasibility neural network is not implemented for this case study.

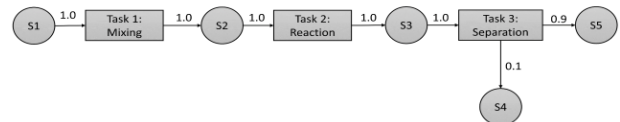

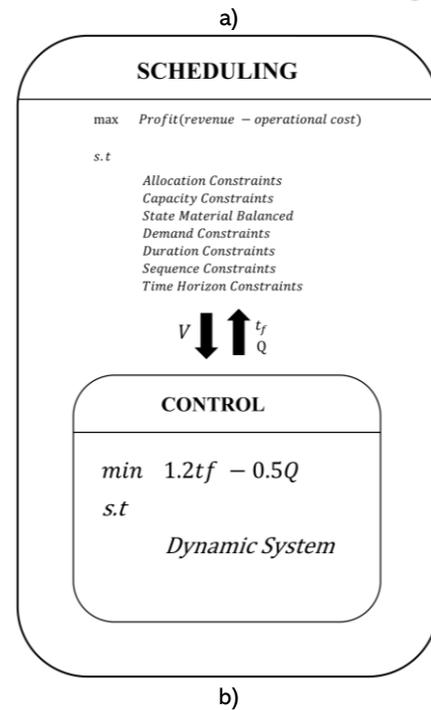

a)

b)

Figure 5.1 a) State task Network representation of the process. b) Communication between the scheduling and control levels

$$\mathbf{y} = \frac{1}{1 + e^{-(\boldsymbol{\omega}^T \mathbf{z}_{l-1} + \mathbf{b})}} \quad (5.3)$$

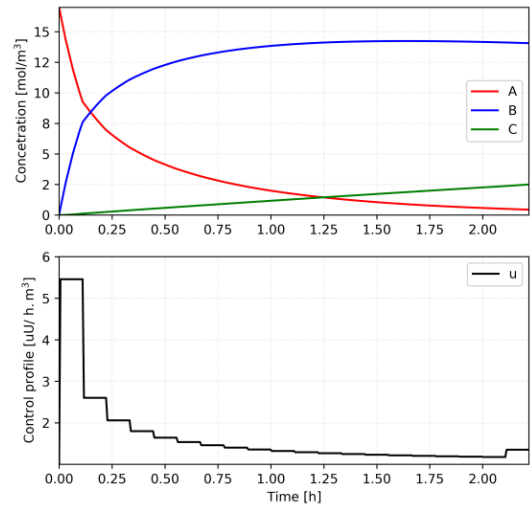

Figure 5.2 Batch reactor optimal control solution example

Note that both output variables must always be zero if the volume is 0. However, due to approximation errors, the neural networks return values close to 0. To overcome this, the processing time and utility consumption is multiplied by the respective binary task assignment variable. Then, the bilinear term is reformulated and replaced using the linking constraints and intermediate variables proposed in Section 3.3 to avoid nonlinearities.

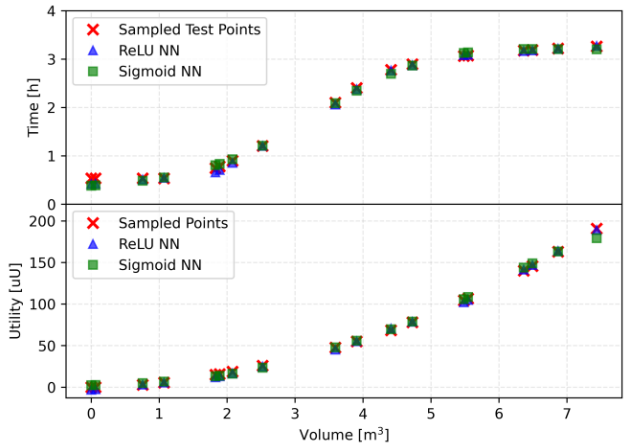

Figure 5.3 Test set predictions

The integrated scheduling and control problem is evaluated for a 12-hour time horizon with five events. Each event requires a solution from the control level. For this time horizon, the optimum number of events is five since adding an extra event does not improve the optimal value.

## 5.2 Results and Discussion

The schedules obtained for the monolithic, NN-MILP, and NN-MINLP Integration approaches can be seen in Figure 5.4. The values in Figures 5.4 b and c are the results obtained from the neural networks; therefore, they are subject to approximation error. These schedules were simulated using the dynamic system, and the actual values obtained are summarised in Table 2. All the solutions produced three batches within the given time horizon. Regarding computational time, the NN-MILP Integration vastly outperforms the other approaches by solving the optimisation problem in 0.2 s. Furthermore, despite the monolithic and NN-MINLP Integration being MINLPs, the former one is solved ten times faster than the monolithic approach. This is expected since the NN-MINLP Integration involves two complex nonlinear expressions (one for each neural network output). At the same time, the monolithic approach needs to solve the entire discretised dynamic system at each iteration (359 constraints).

Table 2. Results summary for Case Study 1 using five events, without considering training time. Actual performance obtained when simulated with the dynamic system.

|  | Monolithic | NN-MILP Integration | NN-MINLP Integration |
|---|---|---|---|
| Computational time (s) | 70.8 | 0.2 | 3.6 |
| Profit (USD) | 280.1 | 273.5 | 268.9 |
| Utility (Utility units [Uu]) | 81.5 | 71.0 | 67.3 |
| Production (m$^3$) | 6.8 | 6.5 | 6.3 |

As shown in Table 2, the monolithic approach is not able to capture the control level's objective (optimal processing time and utility consumption trade-off). The monolithic approach only pursues the scheduling objective, which seeks to produce more and faster regardless of utility consumption. This can be seen by the amount of utility consumed and the production obtained. For example, the monolithic approach produces 6.8 m$^3$ with 81.5 uU with shorter processing times. On the other hand, the NN-MILP-integration produces 6.5m$^3$ in more time but consumes 13% less utility. Therefore, the production reduction is compensated with less utility consumption. Since a local solver is used for the MINLPs, the lower profit obtained from the NN-MINLP integration could be considered a local optimum. Nevertheless, MILP solvers (Gurobi) implement heuristics to help them escape from the local minimum.

The error of the solution due to the neural network approximation could not be quantified by comparing the value of the objective function since the solution from the monolithic approach is inherently different. However, the dynamic optimisation problem is solved separately using the processing time and utility obtained from the neural network integrated approaches to calculate the actual volume that would be produced. Table 3 summarises the relative error obtained for each reactor batch size. As expected, the NN-MINLP Integration has larger errors due to the sigmoid neural network underperformance in the test set. Nevertheless, the performance could be improved by using more data during the training if it is available or by modifying the neural network structure. However, the same neural network architecture is used for this case study to enable a fair comparison.

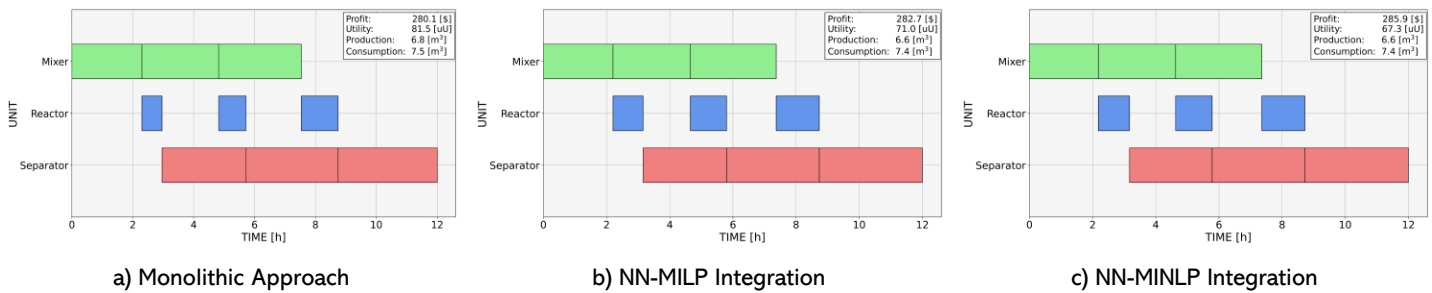

a) Monolithic Approach   b) NN-MILP Integration   c) NN-MINLP Integration

Figure 5.4 Schedules calculated for Case Study 1 using five events

Table 3. Batch sizes' relative errors due to neural network approximation for Case Study 1 with five events.

|  | NN-MILP | NN-MINLP |
|---|---|---|
| Batch 1 relative error (%) | 4.5% | 6.5% |
| Batch 2 relative error (%) | 2.4% | 5.6% |
| Batch 3 relative error (%) | 1.5% | 4.0% |
| Average relative error (%) | 2.8% | 5.4% |

The number of events dictates the size of the scheduling problem. Each event requires an extra neural network for the NN-MILP and NN-MINLP approaches. The ReLU neural network is translated to 40 linear constraints, while the sigmoid neural network is reduced to two nonlinear constraints. On the other hand, an additional dynamic system is added for the monolithic approach (359 constraints). Figure 5.5 shows the performance of the three methods against the problem size. The upper figure shows the optimisation time. For the NN-MILP Integration, adding an event marginally impacts the computational time and is always solved in less than one second. This negligible impact reflects the excellent performance of well-developed MILP solvers for problems within this scale. Moreover, MILP solvers like Gurobi apply heuristics to solve problems faster, which might be the reason for the slight impact and fluctuations.

The bottom figure shows the total time considering the neural network training. However, it is important to notice that the training is performed only once. As can be seen, it is not only worth training a neural network for the smallest case. When considering the training time in the five events scenario, the NN-MILP integration solves the problem 4 times faster than the monolithic approach and 1.2 times faster than the NN-MINLP approach.

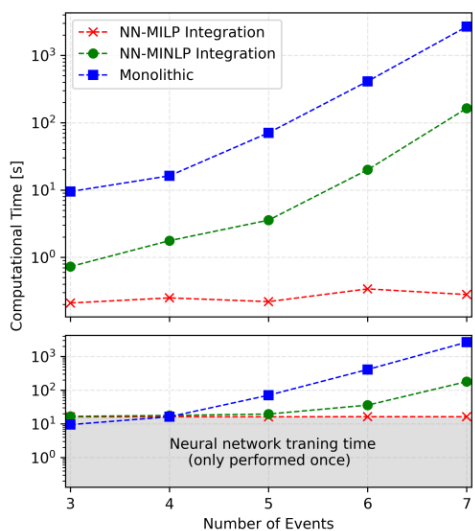

Figure 5.5 Computational time and number of products. Upper figure: Optimisation time, Bottom Figure: Optimisation and neural network training time.

The monolithic and NN-MINLP Integration larger times reflect the computational burden of solving decomposition algorithms within MINLP solvers. Each nonlinear expression from the sigmoid neural network contains 30 exponential terms. This complex expression is responsible for the computational burden of the NN-MINLP Integration. However, the NN-MINLP Integration is still solved faster than the monolithic approach due to the considerably smaller number of constraints added. In literature, shallower neural networks (1 hidden layer) were implemented to avoid these complex expressions. Overall, due to the linear nature and ability to retain the control level's objective, the MILP integration finds the optimal bilevel solution faster.

## 6. Case Study 2

### 6.1 Problem Description

This case study assesses the scheduling of a multiproduct batch reactor. Consider the enzymatic reactions described in Eqs. 6.1. The aim is to produce components $P^A$ and $P^B$ using the same substrate S. Two different enzymes are required ($E_0^A, E_0^B$) for this task. $E_1^A$ and $E_1^B$ are intermediates which represent the occupied enzymes.

$$\begin{aligned} E_0^A + S &\leftrightarrows E_1^A \\ E_1^A &\to E_0^A + P^A \\ E_0^B + S &\leftrightarrows E_1^B \\ E_1^B &\to E_0^B + P^B \end{aligned} \quad (6.1)$$

The desired product is a mix of components $P^A$ and $P^B$, which final concentration depends on the following stages requirement of $P^A$, as can be seen in Table 4. Table 4 also indicates when the products can start to be produced and their due times.

Table 4. Product profile and time restrictions for Case study 2.

| Product | Release Time | Due Time | $P^A_{Demand}$ [g/L] |
|---|---|---|---|
| A | 2 a.m. | 10 a.m. | 0.6 |
| B | 5 a.m. | 21 p.m | 0.75 |
| C | 4 a.m. | 15 p.m | 0.70 |
| D | 0 a.m. | 10 a.m. | 0.64 |
| E | 0 a.m. | 5 a.m. | 0.68 |
| F | 8 a.m. | 15 p.m | 0.74 |
| G | 9 a.m. | 15 p.m | 0.72 |

Overall, the control level calculates the minimum processing time required to satisfy $P^A_{Demand}$ and only has one degree of freedom ($t_{proc}$). On the other hand, the scheduling level pursues the minimisation of the operational cost, which is expressed in terms of operational time (total makespan and delays) and raw material cost (enzymes and substrate). Therefore, the objective of the control level is also contained within the scheduling objective. The total makespan is directly related to the processing time calculated from the control level. This allows quantifying the error due to the neural network approximations. The complete bilevel optimisation problem is described in Figure 6.1, and its notation table can be found in Appendix C. Note that $P^B(t=t_{proc})$ is not a degree of freedom, and its value depends on its kinetics and the minimum processing time required for satisfying $P^A_{Demand}$.

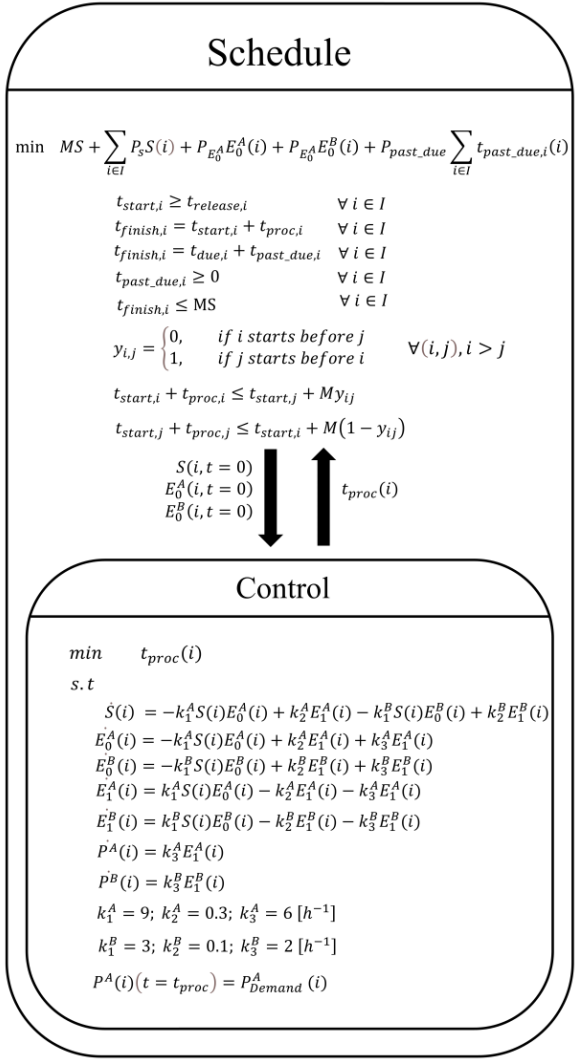

Figure 6.1 Bilevel formulation for Case Study 2

Figure 6.1 also depicts how both levels communicate and the linking variables. The inputs for the neural network are $E^A_0$, $E^B_0$ and S, while the output is the processing time ($t_{proc}$).

A regular mesh of sampling points is built within the bounds of the input variables and $P^A_{Demand}$. The variables bound can be found in Table 5. Since not all values from the sampling points have a feasible solution, infeasibilities are also mapped.

Table 5. Linking Variable bounds.

| Variable | Lower bound [g/L] | Upper bound [g/L] |
|---|---|---|
| S | 0.8 | 1.2 |
| $E^A_0$ | 0.05 | 0.15 |
| $E^B_0$ | 0.01 | 0.1 |
| $P^A_{Demand}$ | 0.6 | 0.8 |

### 6.2 Results and Discussion

The state and feasibility neural networks are implemented in this case. The neural networks' parameters can be seen in Table 6, and their learning curves can be found in Appendix A in Figure A.3. The NN-MILP Integration is benchmarked with the NN-MINLP Integration and monolithic approaches. The ReLU feasibility neural network is used in both cases (NN-MILP and NN-MINLP) to focus on the sigmoid states neural network's impact and avoid neural network classifiers.

Table 6 Neural Network parameters for Case Study 2

| Neural Network | Architecture | Test set performance |
|---|---|---|
| ReLU State | 16:8 | MSE:0.0036 |
| ReLU Feasibility | 16:8 | Accuracy: 98.5% |
| Sigmoid States | 16:8 | MSE = 0.0072 |

To evaluate performance against problem size, the integrated scheduling and control problem is initially solved for two products and scaled up to seven by adding one product at a time. Adding a product requires the upper level to solve another instance of the lower level. As seen in Figure 6.2, the rocketed increase in the number of constraints for the monolithic approach is caused by adding 575 constraints arising from the control level's discretisation. For the NN-MILP Integration, each neural network is translated to 96 linear constraints. For the NN-MINLP integration, 96 linear constraints are added from the ReLU feasibility neural network, while the sigmoid neural adds one large and complex nonlinear constraint.

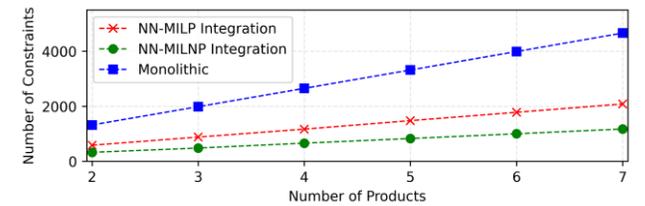

Figure 6.2 Number of constraints added per product

The increasing trend for the solution's error is caused by the neural networks' training error accumulation, as shown in Figure 6.3. The initial decreasing trend for the NN-MILP error is caused by error cancellation due to overestimation and underestimation of the processing time for each product. Overall, the relative error is lower than 6%, which follows the scale of errors reported in Case study 1 and the correlation to the neural network training error. Similar error magnitudes could be obtained from less accurate differential equations discretisation schemes when trying to keep the tractability of MINLPs.

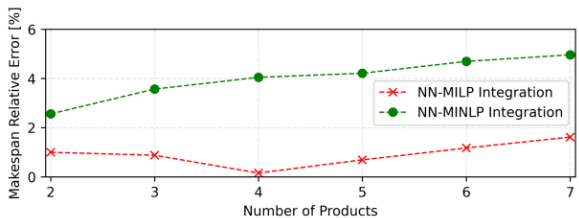

Figure 6.3 Makespan relative error

Figure 6.4 shows the computational time comparison between the three methods. Analogous to the results obtained in the previous case study, there is a substantial reduction in the computational time for the NN-MILP Integration. The NN-MILP integration solves all the cases in less than one second and is slightly impacted by the problem size. Nevertheless, the NN-MINLP outperforms the NN-MILP for very small problems (two products scenario), but its computational time exponentially escalates when adding more products. This increase is caused by the sigmoid neural network, which is formulated as an expression of 144 exponential terms. As a result, the NN-MINLP is solved slower than the monolithic approach for middle-sized problems. Finally, the monolithic approach computational time consistently increases and underperforms the NN-MINLP for large problems. For this case, training a neural network is worth above five products, and it avoids exponential time growth for 6 and 7 products.

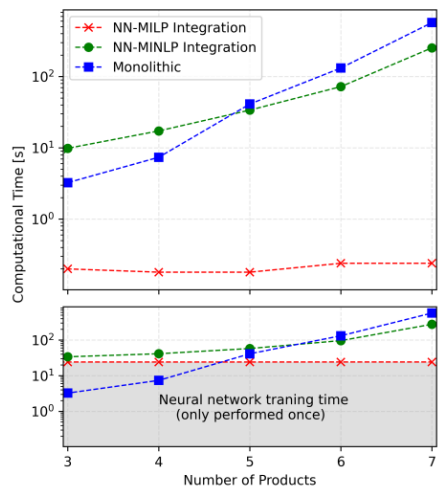

Figure 6.4 Computational time and number of products. Upper figure: Optimisation time, Bottom Figure: Optimisation and neural network training time

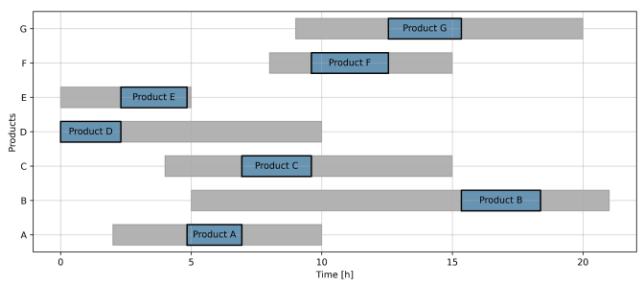

a) Monolithic

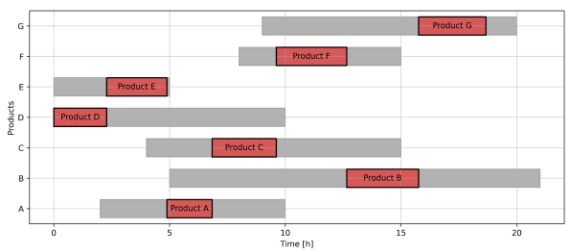

b) NN-MILP Integration

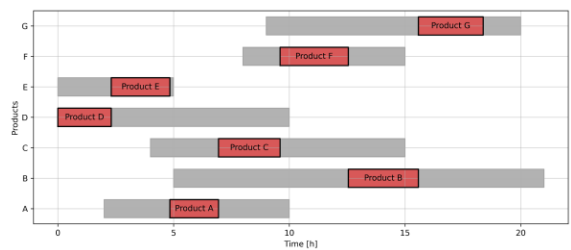

d) Actual schedule with NN-MILP order

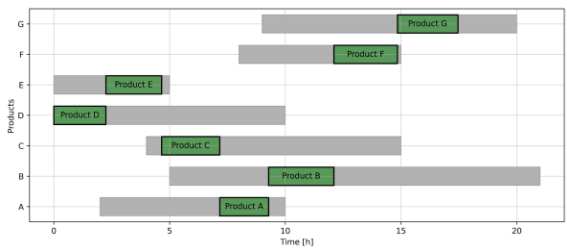

c) NN-MINLP Integration

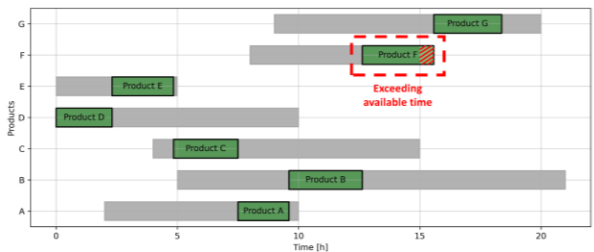

e) Actual schedule with NN-MINLP order

Figure 6.5 Batch reactor schedules result.

Finally, the schedules for the seven products can be seen in Figure 6.5. The grey bars represent the available time window to perform a task, while the coloured bars represent the product processing time. Each methodology obtained a different product order, as seen in Figures 6.5a to 6.5c. Figure 6.5d and 6.5e illustrates the schedules obtained using the order from the NN-MILP and NN-MINLP integration with the real processing time for each product. As shown in Figure 6.5d, the NN-MILP order respects the available time for each product and obtains an objective function value of 91.87 USD. This value is the same as the monolithic approach. On the other hand, the NN-MINLP order exceeds the available time for product E by 0.57 h, as shown in Figure 6.5e. Since delays are allowed but penalised in the objective function, the NN-MINLP finds a suboptimal solution (92.41 USD). This non-optimal solution is directly related to the sigmoid neural network's larger error, which underestimates the processing times. Therefore, although satisfying control level constraints, upper-level constraints could be violated, or suboptimal solutions can be found due to approximation errors.

## 7. Conclusion

To account for integrated decision-making of scheduling and control in optimisation problems, the literature intends either to reformulate it as a single-level (monolithic) or as a bilevel (hierarchical) problem. However, both methods are hindered by the mathematical complexity of combining the high-dimensional NLP derived from the control level with the discrete problem from the scheduling. Overall, we addressed how to seamlessly solve the challenging bilevel scheduling and control problem by taking advantage of smart machine learning and optimisation frameworks. We proposed a single-level reduction by replacing the control level with a ReLU neural network. The neural network was trained with optimal control level solutions for different scheduling decisions and formulated as a MILP via big-M reformulation. In terms of computational performance, this MILP single-level formulation vastly outperformed the monolithic approach and the common bilevel nonlinear optimisation with embedded neural networks. Since the autonomy of the control level was preserved, the presented method could find the actual bilevel solution. Furthermore, it was shown how shallow neural networks are effortless to handle by state-of-the-art MILP solvers and barely impacted the computational time. On the other hand, nonlinearities from the other two methods transformed the problems into non-trivial MINLPs, which are inherently mathematically more complex than MILPs. It was also shown how the error associated with the neural network approximations could lead to suboptimal or infeasible solutions. Future work could take advantage of this single-level reformulation to perform flexibility analyses of the integrated scheduling and control. Furthermore, robust optimisation could be conducted considering the uncertainty caused by the neural network approximation error.

# Appendix A

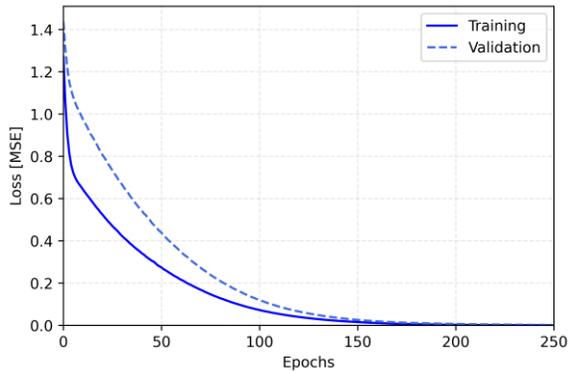

a)

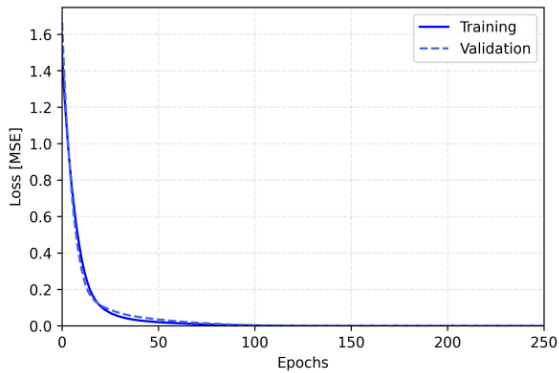

b)

Figure A.1. Learning curves for the neural networks used in the toy problem. a) Aligned objectives b) Adversarial objectives.

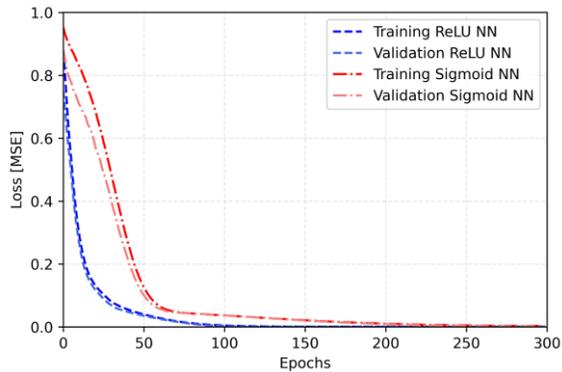

Figure A.2 Learning curves for the neural networks used in Case Study 1.

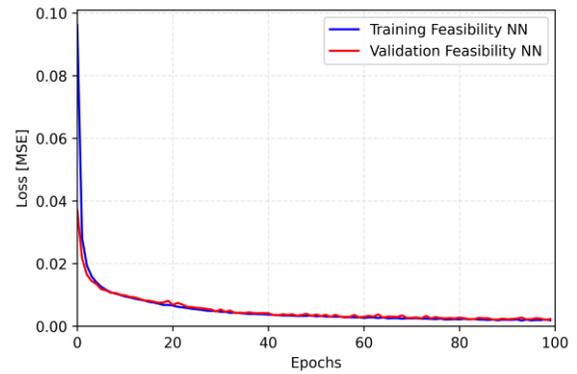

a)

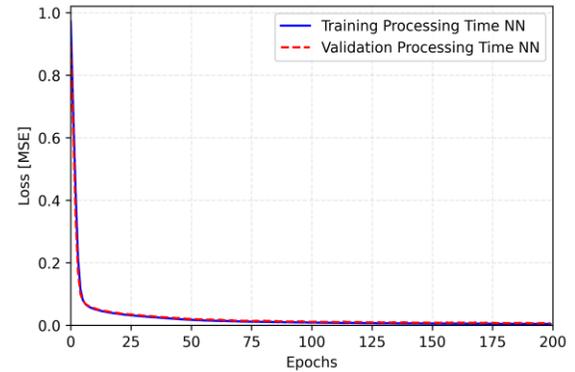

b)

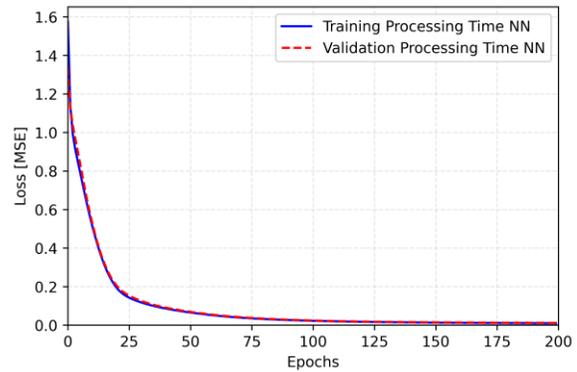

c)

Figure A.3. Learning curves for the neural networks used in Case Study 2.

## 9. Appendix B

The scheduling formulation developed by Ierapetritou & Floudas (1998) was adapted for Case Study 1. The same notation from the original paper is maintained for clarity. According to the process network in Figure 5.1. The process has three units (mixer, reactor, and separator) indexed by j. Three tasks (mixing, reaction, and separation) are indexed by i, and five states (s1, s2, s3, s4 and s5) represent the material leaving and entering each unit. The mathematical formulation is described below, as well as the parameters required.

Sets

I = tasks

$I_j$ = tasks performed in unit j

$I_s$ = tasks that produce or consume state s

J = units

$J_i$ = units capable of executing task i

N = total number of events

S = states

Parameters

$V_{ij}^{min}$ = minimum material amount to be processed by task i at unit j

$V_{ij}^{max}$ = maximum material amount to be processed by task i at unit j

$\rho_{si}^p, \rho_{si}^c$ = fraction of state s produced or consumed by task i, respectively

H = time horizon

Variables

wv(i,n) = binary variable that indicates the assignment of task i at event n

yv(j,n) = binary variable that indicates the assignment of unit j at event n

d(s,n) = total amount of state s sent to the market during event n

ST(s,n) = total amount of state s during event n

$T^s(i,j,n)$ = starting time for task i in unit j during event j

$T^f(i,j,n)$ = end time for task i in unit j during event j

$t_p(i,j,n)$ = processing time of task i at unit j during event n

V(i,j,n) = amount of material processed by task i at unit j during event n

Q(i,j,n) = utility consumed by task i at unit j during event n

$$\max \sum_{s \in S} \sum_{n \in N} d(s,n) \times sell\_price(s) - \sum_{s \in S} d(s,0) \times buy\_price(s) - \sum_{i \in I} \sum_{j \in J} \sum_{n \in N} Q(i,j,n) \quad \text{Eq. B0}$$

$$\sum_{i \in I_j} wv(i,n) = yv(j,n) \; \forall j \in J, n \in N \quad \text{Eq.B1}$$

$$V_{ij}^{min} \leq V(i,j,n) \leq V_{ij}^{max} \quad \text{Eq.B2}$$

$$ST(s,n) = ST(s,n-1) - d(s,n) + \sum_{i \in I_s} \rho_{si}^p \sum_{j \in J_i} V(i,j,n-1) - \sum_{i \in I_s} \rho_{si}^c \sum_{j \in J_i} V(i,j,n-1) \quad \text{Eq.B3}$$
$$\forall s \in S, \; n \in N$$

$T^f(i,j,n) = T^s(i,j,n) + t_f(i,j,n) \quad \forall i \in I, j \in J_i, n \in N$ Eq.B4

$T^s(i,j,n+1) \geq T^f(i,j,n) - H(2 - wn(i,n) - yv(j,n)),$
$\forall j \in J, n \in N, n \neq N$ Eq.B5

$T^s(i,j,n+1) \geq T^s(i,j,n)$
$\forall j \in J, n \in N, n \neq N$ Eq.B6

$T^f(i,j,n+1) \geq T^f(i,j,n)$
$\forall j \in J, n \in N, n \neq N$ Eq.B7

$T^s(i,j,n+1) \geq T^f(i',j,n) - H(2 - wn(i',n) - yv(j,n))$
$\forall j \in J, i \in I_j, i' \in I_{j'}, i \neq i', n \in N, n \neq N$ Eq.B7

$T^s(i,j,n+1) \geq T^f(i',j',n) - H(2 - wn(i',n) - yv(j',n))$
$\forall j, j' \in J, i \in I_j, i' \in I_{j'}, i \neq i', n \in N, n \neq N$ Eq.B9

$T^s(i,j,n+1) \geq \sum_{n' \in N, n' \leq n} \sum_{i' \in I_j} (T^f(i',j,n) - T^s(i',j,n')),$
$\forall j \in J, n \in N, n \neq N$ Eq.B10

$T^f(i,j,n) \leq H$
$T^s(i,j,n) \leq H$
$\forall j \in J, n \in N, n \neq N$ Eq.B11

| State | Buying Price (USD) | Selling Price (USD) |
|---|---|---|
| S1 | 60 | 0 |
| S2 | 0 | 0 |
| S3 | 0 | 0 |
| S4 | 0 | 0 |
| S5 | 0 | 120 |

| Unit | Max Capacity ($V^{max}$) $m^3$ | Max Capacity ($V^{min}$) $m^3$ |
|---|---|---|
| Mixer | 5 | 2 |
| Reactor | 5 | 2 |
| Separator | 5 | 2 |

The processing time and utilities for the mixer and separator are calculated as described below.

$t_f(mixing, mixer, n) = 1 \times V(mixing, mixer, n),$
$\forall n \in N$ Eq.B12

$t_f(separation, separator, n) = 1.2 \times V(separation, separator, n),$
$\forall n \in N$ Eq.B13

$Q(separation, separator, n) = Q(mixing, mixer, n) = 0$
$\forall n \in N$ Eq.B14

## 10. Appendix C

The following notation is used in Figure 6.1 to describe the mathematical formulation of Case Study 2.

$i = tasks, I = set\ of\ tasks$
$MS = total\ makespan$
$t_{release,i} = release\ time\ of\ task\ i$
$t_{due,i} = due\ time\ of\ task\ i$
$t_{start,i} = starting\ time\ of\ task\ i$
$t_{proc,i} = processing\ time\ of\ task\ i$
$t_{finish,i} = finishing\ time\ of\ task\ i$
$t_{past\_due,i} = past\ due\ time\ of\ task\ i$
$M = bigM\ constant$
$P_s, P_{E_0^A}, P_{E_0^B} = Prices\ of\ substrate\ and\ enzymes$
$P_{past\_due} = Economic\ Penalisation\ due\ to\ delays$